\documentstyle[epsf]{elsart}
\begin{document}
\def\myfig #1#2#3#4{\par
\epsfxsize=#1 cm
\moveright #2cm
\vbox{\epsfbox{#3}}
{\noindent Figure~#4 }\vskip .3cm }
\def\lg{\rm log}
\def\bh{black hole~}
\def\ew{equivalent width~}
\def\ad{accretion~disk~}
\def\ads{accretion~disks~}
\def\el {emission-line~}
\def\els {emission-lines~}
\def\bhs{black~holes~}
\def\ar{accretion~rate}
\def\ip{ionization-parameter~} 
\def\ed {Eddington~}
\def\ers{{\rm erg/sec}}
\def\ms{M_{\odot}}
\def\et{{\it et al.~}}
\def\sw{\rm Schwarzshild~}
\def\bb{black body~}
\def\be{Baldwin effect~}
\def\cmq{cm$^{-3}$}
\def\kms{km s$^{-1}$}
\def\ltsim{\raisebox{-.5ex}{$\;\stackrel{<}{\sim}\;$}}
\def\gtsim{\raisebox{-.5ex}{$\;\stackrel{>}{\sim}\;$}}
\def\apj{ApJ}
\def\apjl{ApJL}
\def\apjs{ApJ Supp.}
\def\mnras{MNRAS}
\def\aap{A\&A}

\begin{frontmatter}

\title{Black Holes and Host Galaxies of  NLS1s}

\author[]{A. Wandel}
\address[]{Racah Institute of Physics, The Hebrew University of Jerusalem, 
Israel}

\begin{abstract}
Recently, reliable mass estimates for the central black holes in AGN became 
feasible due to emission-line reverberation techniques. Using this method 
as a calibrator, it is possible 
to determine black hole masses for a wide range of AGN,
in particular NLS1s. Do NLS1s have smaller black holes than ordinary Seyfert
1 galaxies? Are their black holes smaller compared to the sizes of their host
galaxies? Do they have larger $L/M$ ratios? 
Do NLS1s have hotter accretion disks? I confront these questions with accretion
disk theory and with the data, showing that the above may well be the case.
\end{abstract}

\begin{keyword}
galaxies: active; quasars: general; quasars: emission lines; accretion disks;
black holes
\end{keyword}

\end{frontmatter}



\section{The spectrum-\bh mass relation for \ads }

In the \ad paradigm for the power source of an AGN, the continuum luminosity
and spectral temperature are related to the \bh mass by the relation

\begin{equation}
\label {equ:elmr}
L_{46}\sim (E/10~{\rm eV})^{4}M_8^{-2}(R/5 R_{\rm s})^{2},
\end{equation}
where  $E$ is the average photon
energy, $R_{\rm s}=2GM/c^2\approx 3\times 10^{13} M_8~$cm is the \sw radius, and
$L_{46}$ is the bolometric luminosity.
The temperature in the \bb part of the disk is
\begin{equation}
\label {equ:tr}
T(R) \approx \left (  {3GM\dot M\over 8\pi\sigma R^3} \right )^{1/4} 
\approx 6\times 10^5 \left (  {\dot m\over M_8} \right ) ^{1/4} r^{-3/4}~{\rm K},
\end{equation}
\noindent
where $r=R/R_{\rm s}$ and  
$\dot m=\dot M/\dot M_{\rm Edd}\approx 2(\dot M/\ms~{\rm yr}^{-1}) M_8^{-1} (\epsilon/0.1) 
^{-1}$
is the \ar~ in units of the \ed \ar~ ($\epsilon$ being the efficiency), so that $\dot m =1$ at the \ed limit.
When the \ar~ approaches the \ed rate, the thin disk solution in the
inner region is probably not valid, and it has to be replaced
by a hot disk solution (e.g. Wandel \& Liang 1991).
The spectrum of a multi-\bb \ad is given by integrating the local \bb spectrum
over the entire disk, 
$
L_\nu  \approx \int_{R_{\rm t}}^{R_{\rm out}} 2\pi R B_\nu [T(R) ] dR
$,
%
where $B_\nu(T)$ is the Planck function and
$R_{\rm t}$ is the transition radius from the \bb region to the inner 
optically thin region.
%
%
If the disk is radially extended ($R_{\rm out}/R_{\rm t}\gg 1$), the spectrum is almost flat 
($\sim \nu^{1/3}$)
and cuts off beyond $h\nu_{\rm co}\approx 3kT(R_{\rm t})$, 
approximately
\begin{equation}
\label {equ:lker}
L_\nu  \approx A \left ({\nu\over \nu_{\rm co}}\right )^{1/3} \exp
\left (-{\nu\over \nu_{\rm co}}\right ),
\end{equation}
where $A$ and $\nu_{\rm co}$ are the normalization and cutoff frequency.
For a Kerr black hole, Malkan (1990) finds
$\nu_{\rm co} = (2.9\times 10^{15}~{\rm Hz}) \dot M_{0.1}^{1/4}M_8^{-1/2}$,
where $\dot M_{0.1}=\dot M/0.1 \ms~{\rm yr}^{-1}$. This can be written as
\begin{equation}
\label {equ:eco}
h\nu_{\rm co} = (6~{\rm eV}) \dot m^{1/4}M_8^{-1/4}= (20~{\rm eV}) L_{46}^{1/4}M_8^{-1/2},
\end{equation}
where $L_{46}$ is the {\it observed} luminosity, and we have  included 
a bolometric correction of 10. 
If the \bb region extends down to a radius $R_{\rm t}$, and the EUV cutoff energy is 
$E_{\rm co}$,
then  
\begin{equation}
\label {equ:mte}
M_8\approx 2.5 (E_{\rm co} /10~{\rm eV})^{-2}L_{46}^{1/2} (R_{\rm t}/5 R_{\rm s})^{-1}.
\end{equation}
When the \bb regime extends close to the inner disk edge, 
the spectrum for the \sw case peaks at the photon energy
$E\gtsim (17~{\rm eV})L_{46}^{1/4}M_8^{-1/2}$.
For a Kerr \bh (eq. \ref{equ:eco})
$E\gtsim (20~{\rm eV})L_{46}^{1/4}M_8^{-1/2} $.
Plotting the cutoff frequency versus the luminosity, one can infer
the mass. Alternatively, if the mass can be estimated independently, it is
possible to estimate the cutoff frequency (Fig.~1;
cf. Wandel \& Petrosian 1988).

\myfig {12} {1} {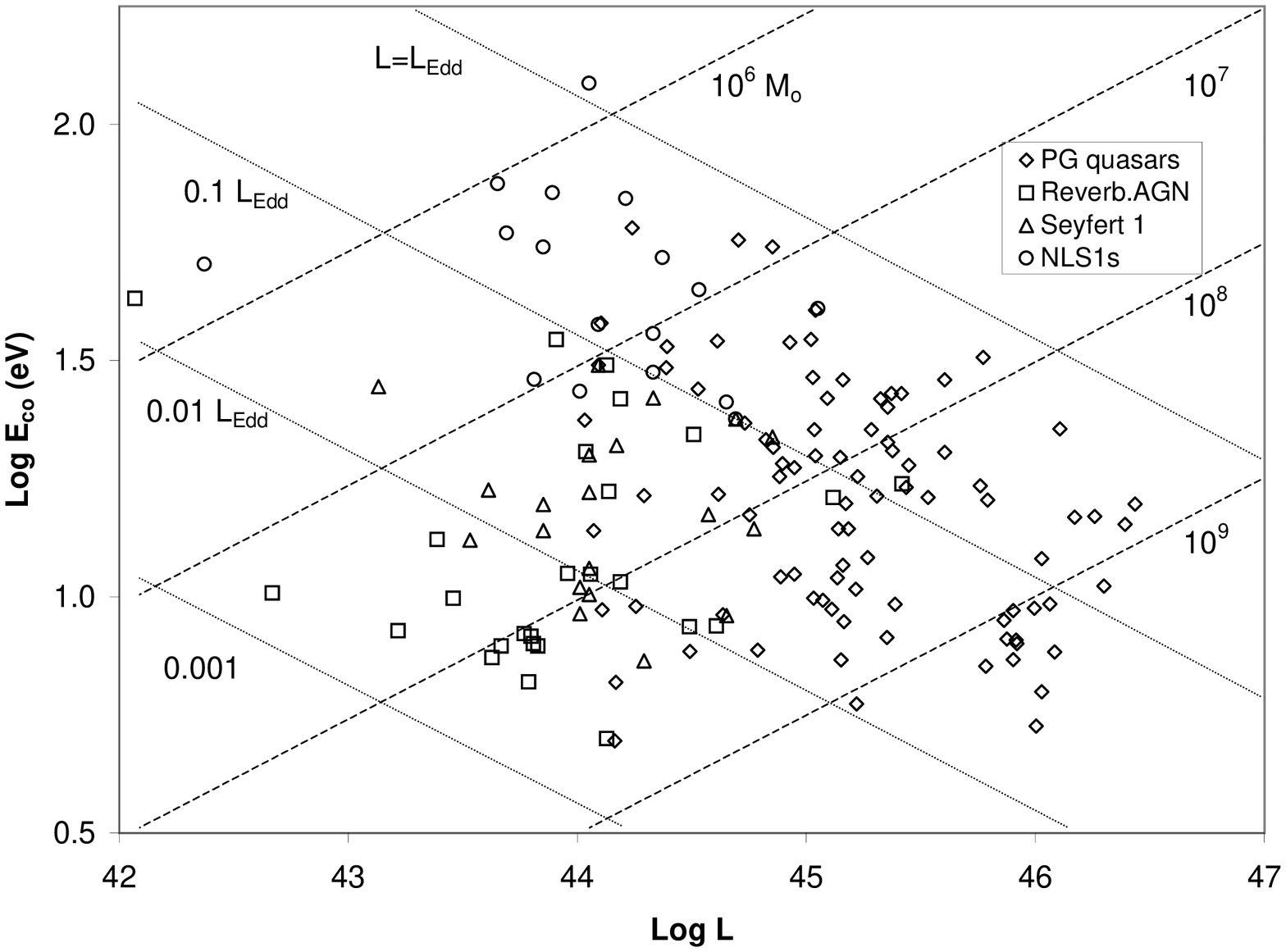}
{1. The peak (or cutoff) energy for an \ad spectrum versus the monochromatic 
luminosity at
5100~\AA. Diagonal dashed lines indicate constant \bh masses, and diagonal 
dotted lines indicate the \ed ratio.
Diamonds indicate PG quasars (Boroson \& Green 1992), squares---Seyfert 1
galaxies, triangles---AGN with BLR reverberation data, circles---NLS1s
(Boller, Brandt \& Fink 1996).}
\section{Black Hole Masses}
It is possible to estimate the \bh mass from the broad emission-line profile,
assuming the velocity width is induced by a Keplerian velocity dispersion.
When the broad line region size is estimated by reverberation mapping,
this technique is particularly reliable. Wandel, Peterson \& Malkan (1999) have 
used a sample of AGN with reverberation data to calibrate the mass estimate
obtained from photoionization models, finding the relation
\begin{equation}
\label {equ:mph}
M_8\approx 0.4 \left ({L_{46}\over Un_{10}}\right )^{1/2} v_3^2,
\end{equation}
where $U$ is the ionization parameter, $n_{10}$ is the density in units
of 10$^{10}$~\cmq, and $v_3$ is the H$\beta$ FWHM in units of $10^3$~\kms .
Using this calibrated relation, it is possible to estimate \bh masses
for large samples,
even without reverberation data (Wandel, Malkan \& Peterson, in preparation).
Combining equations \ref{equ:mph} and \ref{equ:eco}, 
we have 

\begin{equation}
\label {equ:ecov}
E_{\rm co}\approx (40~{\rm eV}) (Un_{10})^{-1/4} v_3^{-1},
\end{equation}
which is independent of luminosity and only weakly dependent on the unknown
parameters $U$ and $n$. Using eq. \ref{equ:ecov}
with $Un_{10}$=1, we find the mass and cutoff energy distribution
for a large sample of quasars, Seyfert 1 galaxies and NLS1s (Fig.~1).
Note that $E_{\rm co}$ is anticorrelated with the mass, and in particular
that different AGN categories group in different regions of the
$L$-$E_{\rm co}$ plane: quasars have more massive \bhs and low cutoff energies,
Seyfert~1 galaxies have intermediate \bh masses ($10^7$--$10^8\ms$), and
NLS1s have low \bh masses ($10^6$--$10^7\ms$) and high cutoff energies.
This is consistent with the large soft X-ray excesses observed for many of
the NLS1s. 
%
The diagonal dotted lines in Fig.~1 give the $L/L_{\rm Edd}$ ratio. It appears 
that quasars tend to be in the 
0.01--1 range, Seyferts in the 0.001--0.1 range, 
and NLS1s are all near \ed with 
$L/L_{\rm Edd}\approx$~0.1--1. 

\section{The BH-Bulge relation}
Compact dark masses, probably massive \bhs (MBHs), have been detected in the 
cores of 
many normal galaxies using stellar dynamics (Kormendy \& Richstone 1995).
The MBH mass appears to correlate with the galactic
bulge luminosity, with the MBH 
 having about one percent of the mass of the spheroidal bulge
(Magorrian \et 1998, Richstone \et 1998). 

The question of whether AGN, and NLS1s in particular, follow a similar \bh-bulge 
relation is a very interesting one, as it may shed light 
on the connection between the host galaxy and the active nucleus.

Wandel \& Mushotzky (1986) have found an excellent correlation between the 
virial mass included within the narrow line region (of order tens to 
hundreds of pc from the center) and the \bh~mass estimated from X-ray variability 
in a sample of Seyfert 1 galaxies, while a group of Seyfert~2 galaxies 
systematically deviated from this relation.
A \bh-bulge relation similar to that of normal galaxies 
has been reported between the MBH of bright PG quasars and their 
host galaxies (Laor 1998), but the mass of the elliptical host  
(or bulge, for spiral hosts) estimates have large uncertainties. 
A significantly lower \bh to bulge mass ratio has been found for the
17 Seyfert~1 galaxies with reverberation data (Wandel 1999). While the average
is lower than the Magorrian \et relation by a factor of 20, the 
record belongs to the NLS1 galaxy NGC~4051, for which Wandel (1999) finds 
a ratio that is a factor of 200 lower than the Magorrian \et relation.

Apparently, this result suggests an intrinsic difference between the central 
\bhs of Seyfert galaxies and those of normal galaxies.
Actually the difference may be (at least partly) due to a selection effect. 
In angular-resolution limited methods (which are applied for detecting MBHs 
in normal galaxies), the MBH detection limit 
is correlated with bulge luminosity: more luminous
bulges have a higher detection limit because the stellar velocity
dispersion is higher (the Faber-Jackson relation). In order to detect the 
dynamic effect of a MBH, it is necessary to observe closer to the center,
while the most luminous galaxies tend to be at  larger distances. So, for
a given angular resolution, the MBH detection limit is higher.
Another effect could bias the PG quasars. Being  the brightest nearby
quasars, they may represent a subset of massive (and well-fed) \bhs.
Smaller ones would not appear as bright and would not be included in the quasar
host galaxy survey.

The BLR method is not subject to this constraint,
making Seyfert 1 galaxies good candidates for detecting  
low-mass MBHs. 
Seyfert~1 galaxies also have more reliable 
bulge mass estimates; as they are nearer and have a lower nuclear brightness,
their host type and bulge magnitudes can be estimated more easily 
(Whittle 1992). 

Finally, we concentrate on the difference
 between Seyfert 1 galaxies and NLS1s. NGC 4051, the only NLS1 in the sample 
of 17 Seyfert~1 galaxies with reverberation data, 
shows a dramatically lower \bh to bulge mass ratio and higher $L/M$ ratio 
(Wandel 1999). 
Reverberation data on more
NLS1s are required before one can conclude that NLS1s have intrinsically
lower \bh to bulge mass ratios and smaller black holes.


\end{document}